\documentclass{ptephy_v1}

\usepackage{amsmath}
\usepackage{graphics}
\usepackage{url}

\def\edth{{\rlap{$\partial$}\raise0.3em\hbox{$-$}}}
\newcommand{\bea}{\begin{eqnarray}}
\newcommand{\eea}{\end{eqnarray}}

\begin{document}

\title{
Estimate of the radius responsible for 
quasinormal modes in the extreme Kerr limit
and
asymptotic behavior of
\\
the Sasaki--Nakamura transformation}


\author{\name{Hiroyuki Nakano}{1},
\name{Norichika Sago}{2}, 
\name{Takahiro Tanaka}{1,3},
and \name{Takashi Nakamura}{1}}
\address{${}^1$\affil{1}{Department of Physics, Kyoto University, Kyoto 606-8502, Japan}
\\
${}^2$\affil{2}{Faculty of Arts and Science, Kyushu University, Fukuoka 819-0395, Japan}
\\
${}^3$\affil{3}{Yukawa Institute for Theoretical Physics, Kyoto University, Kyoto 606-8502, Japan}
}

\begin{abstract}
The Sasaki--Nakamura transformation gives
a short-ranged potential and a convergent source term for the master equation
of perturbations in the Kerr space-time.
In this paper, we study the asymptotic behavior of
the transformation, and present a new relaxed necessary and sufficient condition
for the transformation to obtain the short-ranged potential
in the assumption that the transformation converges in the far distance.
Also, we discuss the peak location of the potential
which is responsible for quasinormal mode frequencies in tWKB analysis.
Finally, in the extreme Kerr limit, $a/M \to 1$,
where $M$ and $a$ denote the mass and spin parameter of a Kerr black hole,
respectively, we find the peak location of the potential,
$r_{\rm p}/M \lesssim 1 + 1.8 \,(1-a/M)^{1/2}$
by using the new transformation. The uncertainty
of the location is as large as that expected from the equivalence principle.
\end{abstract}

\subjectindex{E31, E02, E01, E38}

\maketitle

\section{Introduction}

The source of GW150914, a gravitational wave (GW) event 
observed by advanced LIGO on September 14, 2015~\cite{Abbott:2016blz},
is considered as a merging binary black hole (BBH)
and the black hole (BH) masses are estimated as 
$36^{+5}_{-4}M_\odot$ and $29^{+4}_{-4}M_\odot$, respectively.
According to Ref.~\cite{TheLIGOScientific:2016htt},
the BH masses are well predicted by 
the recent population synthesis results of Population III
BBHs~\cite{Kinugawa:2014zha,Kinugawa:2015nla,Kinugawa:2016mfs}
(see also Ref.~\cite{Hartwig:2016nde}).

The mass and (non-dimensional) spin of the remnant BH
were estimated as 
$62^{+4}_{-4}M_\odot$ and $0.67^{+0.05}_{-0.07}$~\cite{TheLIGOScientific:2016wfe}
by using the model derived in Refs.~\cite{Healy:2014yta,Ghosh:2015jra}, respectively.
But, the signal from the ringdown phase of GWs, described by
quasinormal modes (QNMs) of a BH, was too weak to test general relativity
(see Refs.~\cite{Berti:2007zu,Nakano:2015uja},
and also Refs.~\cite{Konoplya:2016pmh,Yunes:2016jcc}),
and only a consistency check for the least-damped QNM
has been done in Ref.~\cite{TheLIGOScientific:2016src}.
However, since the expected event rate is
high~\cite{Kinugawa:2014zha,Kinugawa:2015nla,Kinugawa:2016mfs},
and the sensitivities of GW observations will improve,
there will be a good chance to have an event with much higher
signal-to-noise ratio.

To consider QNMs, we use the BH perturbation approach.
The Kerr metric~\cite{Kerr:1963ud}
in the Boyer--Lindquist coordinates is written as
\bea
 ds^2 &=& 
 - \left( 1 - \frac{2 M r}{\Sigma} \right)  dt^2 
 - \frac{4 M a r ~{\rm{sin}^2 \theta} }{\Sigma} dt d\phi 
 + \frac{\Sigma}{\Delta} dr^2
 \cr &&
 + \Sigma d\theta^2
 + \left( r^2 + a^2 + \frac{2 M a^2 r}{\Sigma} \sin^2 \theta \right)
 \sin^2 \theta d\phi^2 \,,
\eea
where $\Sigma = r^2 + a^2 \cos^2 \theta$, $\Delta = r^2 - 2 M r + a^2$,
and $M$ and $a$ denote
the mass and the spin parameter of a Kerr BH, respectively.
The Kerr space-time is the background to calculate BH perturbations.

Perturbations are discussed by using the Teukolsky formalism.
The radial Teukolsky equation~\cite{Teukolsky:1973ha}
for gravitational perturbations in the Kerr space-time
is formally written as
\begin{eqnarray}
 \Delta^2\frac{d}{dr}\frac{1}{\Delta}\frac{dR}{dr}-VR = -T \,,
\label{eq:Teuk}
\end{eqnarray}
where $T$ is the source and the potential $V$ is given by
\begin{eqnarray}
 V = -\frac{K^2}{\Delta}-\frac{2iK\Delta'}{\Delta}+4iK'+\lambda \,,
\label{eq:TeukV}
\end{eqnarray}
with
\begin{eqnarray}
 K = \left(r^2+a^2\right)\,\omega - am \,.
 \label{eq:K}
\end{eqnarray}
The constants $m$ and $\lambda$ in the Teukolsky equation
label the spin-weighted spheroidal harmonics
$Z^{a\omega}_{\ell m}(\theta,\,\phi)$. 
$\lambda$ is the separation constant which depends on $m$ and $a\omega$.
A prime denotes the derivative with respect to $r$.

There are various modifications of the original Teukolsky equation
proposed to improve the behavior
of the potential $V$ and the source term $T$.
For example, in Ref.~\cite{Chandrasekhar:1976zz} (and related references therein),
Chandrasekhar and Detweiler developed various transformations in the 1970s.
In Refs.~\cite{Nakamura:2016gri,Nakano:2016sgf}, we used
the Detweiler potential given in Ref.~\cite{Detweiler:1977gy}
to study QNM frequencies in WKB
analysis~\cite{Mashhoon:1985cya,Schutz:1985zz}.
Sasaki and Nakamura~\cite{Sasaki:1981kj,Sasaki:1981sx,Nakamura:1981kk}
considered a transformation to remove the divergence in the source term
and to obtain the short-ranged potential.
This Sasaki--Nakamura transformation has been generalized
for various spins in Ref.~\cite{Hughes:2000pf}.

In the WKB analysis, the QNM frequencies are calculated by
\bea
(\omega_r + i \omega_i)^2 = {\cal V}(r^*_0) 
-i\left(n+\frac{1}{2}\right) \sqrt{-2 \left. \frac{d^2{\cal V}}{dr^{*2}} \right|_{r^*=r^*_0}} \,,
\label{eq:WKB_QNM}
\eea
with $n=0,\,1,\,2,\, \cdot\cdot\cdot$.
Here, $\omega_{\rm r}$ and $\omega_{\rm i}$ are the real and imaginary parts
of the frequency, respectively,
$r^*_0$ denotes the location where the derivative of the potential
$d{\cal V}/dr^*=0$ in the tortoise coordinate
$r^*$ defined by $dr^*/dr = (r^2+a^2)/\Delta$,
and we focus only on the $n=0$ mode in this paper.
It is noted that $r^*_0$ is complex-valued in general.

In Refs.~\cite{Nakamura:2016gri,Nakano:2016sgf,Nakamura:2016yjl},
we have used the potential ${\cal V}$ with the substitution
of accurate numerical results of
the complex QNM frequencies~\cite{Berti:2005ys}\footnote{
\url{http://www.phy.olemiss.edu/~berti/ringdown/}.}
obtained by the Leaver's method~\cite{Leaver:1985ax}
to determine the above $r^*_0$ of the potential.
In practice, the peak location $r^*_{\rm p}$ of $|{\cal V}|$,
a real-valued radius,
is also used because we have seen a good agreement between 
the real part of $r^*_0$ and $r^*_{\rm p}$~\cite{Nakamura:2016gri}.
Then, we have compared the QNM frequency calculated
by Eq.~\eqref{eq:WKB_QNM} with $r^*_0$ (or $r^*_{\rm p}$)
in the WKB method with that from the numerical result.
The difference provides an error estimation used
to establish the physical picture that the QNM brings 
information around the peak radius.
Here, we implicitly assume that the peak location is relevant
to the generation of the QNM if the estimated error is small.

The analysis of the peak location of the potential in the extreme Kerr limit
has been discussed based on a single form
of the potential ${\cal V}$ in Ref.~\cite{Nakamura:2016yjl}.
Here, we also evaluate the uncertainty 
in the analysis of the peak location
that originates from the fact that
the GWs cannot be localized due to the equivalence principle,
by comparing various forms of the potential.

This paper is organized as follows. In Sect.~\ref{sec:SN},
we briefly review the Sasaki--Nakamura
transformation~\cite{Sasaki:1981kj,Sasaki:1981sx,Nakamura:1981kk}
and present and discuss a new transformation introduced
in Ref.~\cite{Nakamura:2016yjl}.
In Sect.~\ref{sec:eK}, the peak location of the potential
is calculated in the extreme Kerr limit.
The peak location is related to the mass and spin of the Kerr BH
with expected uncertainties.
Section~\ref{sec:dis} is devoted to discussions.
In Appendix~\ref{sec:SNw} we give a brief summary
of WKB analysis for the QNMs.
We use the geometric unit system, where $G=c=1$ in this paper.

\section{The Sasaki--Nakamura equation and its modification}\label{sec:SN}

The Teukolsky equation in Eq.~\eqref{eq:Teuk} has undesired features.
One is that the source term $T$ diverges as $\propto r^{7/2}$
when we consider a test particle falling into a Kerr BH as the source.
Also, the potential $V$ in Eq.~\eqref{eq:TeukV} is a long-ranged one.
To remove these undesired features, 
Sasaki and Nakamura~\cite{Sasaki:1981kj,Sasaki:1981sx,Nakamura:1981kk}
considered a change of variable and potential.
Since we deal with QNMs in this paper, we focus on the
homogeneous version of the Sasaki--Nakamura formalism in the beginning.

Using two functions $\alpha(r)$ and $\beta(r)$ unspecified for the moment,
we introduce various variables as
\begin{eqnarray}
 X &=& \frac{\sqrt{r^2+a^2}}{\Delta}
 \left(\alpha R+\frac{\beta}{\Delta}R'\right) \,,
 \label{eq:XR} \\
 \gamma &=& \alpha \left(\alpha+\frac{\beta'}{\Delta}\right)
 -\frac{\beta}{\Delta}\left(\alpha'+\frac{\beta}{\Delta^2}V\right) \,,
 \label{eq:gamma} \\
 F &=& \frac{\Delta}{r^2+a^2}\frac{\gamma'}{\gamma} \,,
 \label{eq:F} \\
 U_0 &=& V+\frac{\Delta^2}{\beta}
 \left[ \left(2\alpha+\frac{\beta'}{\Delta}\right)'
 -\frac{\gamma'}{\gamma}\left(\alpha+\frac{\beta'}{\Delta}\right)\right] \,,
 \label{eq:U0} \\
 G &=& -\frac{\Delta'}{r^2+a^2}+\frac{r\Delta}{(r^2+a^2)^2} \,,
 \label{eq:G} \\
 U &=& \frac{\Delta U_0}{(r^2+a^2)^2}+G^2+\frac{dG}{dr^*}
 -\frac{\Delta G}{r^2+a^2}\frac{\gamma'}{\gamma} \,.
 \label{eq:U}
\end{eqnarray}
Then, we have a new wave equation for $X$ derived from the Teukolsky equation as 
\begin{equation}
 \frac{d^2X}{dr^{*2}}-F\frac{dX}{dr^*}-UX=0 \,.
\label{eq:diffXeq}
\end{equation}
We specify $\alpha$ and $\beta$ by
\begin{eqnarray}
 \alpha &=& A-\frac{iK}{\Delta} B \,, \\
 \beta &=& \Delta B \,,
\end{eqnarray}
where
\begin{eqnarray}
 A &=& 3iK'+\lambda+\Delta P \,, 
\label{eq:A}
\\
 B &=& -2iK+\Delta'+\Delta Q \,,
\label{eq:B}
\end{eqnarray}
with
\begin{eqnarray}
 P &=& \frac{r^2+a^2}{gh}\left(\left(\frac{g}{r^2+a^2}\right)'h\right)' \,, 
\label{eq:P}
\\
 Q &=& \frac{(r^2+a^2)^2}{g^2h}\left(\frac{g^2h}{(r^2+a^2)^2}\right)' \,,
\label{eq:Q}
\end{eqnarray}
where $g$ and $h$ are free functions.
Using a new variable $Y$ defined by $X=\sqrt{\gamma}~Y$, we have
\begin{eqnarray}
 \frac{d^2Y}{dr^{*2}}+\left( \omega^2-V_{\rm SN} \right) Y =0 \,,
\label{eq:finalSN}
\end{eqnarray}
where
\begin{eqnarray}
 V_{\rm SN}=\omega^2+U
 -\left[ \frac{1}{2}\frac{d}{dr^*}
 \left(\frac{1}{\gamma}\frac{d\gamma}{dr^*}\right)
 -\frac{1}{4\gamma^2}\left(\frac{d\gamma}{dr^*}\right)^2\right] \,.
\label{eq:SN_V}
\end{eqnarray}

In the above Sasaki--Nakamura transformation,
there are two free functions, $g$ and $h$.
The restrictions that guarantee
a short-ranged potential $V_{\rm SN}$ and a convergent source term
have been given by
\begin{eqnarray}
 g = {\rm const}. \,, \quad h={\rm const}. \,,
\label{eq:gh_hor}
\end{eqnarray}
for $r^* \to -\infty$, and
\begin{eqnarray}
 g = {\rm const}. +O(r^{-2}) \,, \quad h={\rm const}. + O(r^{-2}) \,,
\label{eq:gh_inf}
\end{eqnarray}
for $r^* \to +\infty$.
In Refs.~\cite{Sasaki:1981sx, Sasaki:1981kj},
\begin{equation}
 h=1 \,, \quad g=\frac{r^2+a^2}{r^2} \,,
\label{eq:orig_hg}
\end{equation}
are adopted to satisfy the conditions
in Eqs.~\eqref{eq:gh_hor} and \eqref{eq:gh_inf}.

In Ref.~\cite{Nakamura:2016yjl}, however, 
we have introduced a new $g$ defined by
\begin{equation}
 g=\frac{r(r-a)}{(r+a)^2} \quad ({\rm with}~~h=1) \,,
\label{eq:new_g}
\end{equation} 
which turned out to be suitable to discuss the QNM frequencies
in the WKB approximation. 
The new form of the potential derived from this new $g$
has been plotted in Fig.~1 of Ref.~\cite{Nakamura:2016yjl}
up to $a/M=q=0.99999$. From the standpoint that we calculate
the peak of the potential as the location
where the QNM GWs are emitted in the WKB analysis,
while we cannot apply this discussion to the original Sasaki--Nakamura or
the Detweiler potential (used in Refs.~\cite{Nakamura:2016gri,Nakano:2016sgf}),
the new form of the potential with Eq.~\eqref{eq:new_g} allows us
to discuss the extreme Kerr limit.
The choice given in Eq.~\eqref{eq:new_g} shares the same feature
as the original one (Eq.~\eqref{eq:orig_hg}),
in the sense that the Regge--Wheeler potential~\cite{Regge:1957td}
is recovered for $a=0$.

Although $g$ given in Eq.~\eqref{eq:new_g} does not
satisfy the condition~\eqref{eq:gh_inf} for $r^* \to +\infty$
but behaves as ${\rm const}. +O(r^{-1})$,
we have obtained a short-ranged potential,
which motivates us to revisit the asymptotic conditions on $g$ (and $h$).
To investigate the asymptotic behavior of the potential $V_{\rm SN}$
for $r^* \to +\infty$, we assume that the two free functions are expanded as
\bea
 g = g^{[0]} + \frac{g^{[1]}}{r} + \frac{g^{[2]}}{r^2} \,, \quad
 h = h^{[0]} + \frac{h^{[1]}}{r} + \frac{h^{[2]}}{r^2} \,,
\label{eq:mod_gh}
\eea
where $g^{[n]}$ and $h^{[n]}$ ($n=0$, $1$, $2$) are 
$r$-independent coefficients, and $g^{[0]} \neq 0$
and  $h^{[0]} \neq 0$.
For simplicity, we set $M=1$ in the following.

When $h^{[1]}/r$ does not vanish, $\gamma$ given
in Eq.~\eqref{eq:gamma} has $O(r^1)$ terms
which become $O(r^{-1})$ for $F$ in Eq.~\eqref{eq:F}.
Then, we have $O(r^{-1})$ terms in the potential,
which indicates that the potential is long-ranged.
On the other hand, the term $g^{[1]}/r$ derives 
$O(r^0)$ in $\gamma$ defined by Eq.~\eqref{eq:gamma},
and does not contribute to any $O(r^{-1})$ term in the potential.

More precisely, if we choose $h^{[1]}=0$ in Eq.~\eqref{eq:mod_gh},
we find
\bea
 P = \frac{6}{r^2} + \frac{3{\cal G}}{r^3} + O(r^{-4}) \,, \quad
 Q = -\frac{4}{r} - \frac{{\cal G}}{r^2} + O(r^{-3}) \,,
\label{eq:scPQ}
\eea
for $r \to +\infty$, where ${\cal G}=2g^{[1]}/g^{[0]}$.
Although the above asymptotic behavior of $P$ and $Q$ is different
from that presented in Eq.~(A.4) in Ref.~\cite{Sasaki:1981sx}
(cf. $P=6/r^2+(r^{-4})$ and $Q=-4/r+O(r^{-3})$ in Ref.~\cite{Sasaki:1981sx}),
$A$ and $B$ in Eqs.~\eqref{eq:A} and \eqref{eq:B} have the same asymptotic behavior
as given in Eq.~(A.5) of Ref.~\cite{Sasaki:1981sx} and
$\gamma={\rm const}.+O(r^{-1})$.
This fact guarantees $V_{{\rm SN}}$ to be short-ranged.
Namely, $V_{{\rm SN}}=O(r^{-2})$ is achieved
under the less restrictive condition, $h^{[1]}=0$.
It is worth noting that the asymptotic behavior given in Eq.~\eqref{eq:scPQ}
does not depend on the choice of
$g^{[0]}$, $g^{[1]}$, $g^{[2]}$, $h^{[0]}$, or $h^{[2]}$.

As a summary, we conclude that 
the sufficient condition for $r^* \to +\infty$ can be
relaxed from Eq.~\eqref{eq:gh_inf} to
\begin{eqnarray}
 g = {\rm const}. +O(r^{-1}) \,, \quad h={\rm const}. + O(r^{-2}) \,.
\label{eq:gh_inf_new}
\end{eqnarray}
Under the assumption that the two free functions have
the forms of Eq.~\eqref{eq:mod_gh} at $r^* \to \infty$,
we find that $h^{[1]}=0$ is also the necessary condition.
Although we do not discuss here the inhomogeneous version of
the Sasaki--Nakamura formalism, i.e., the source term, in detail,
it is easily found that the transformation under
the above conditions~\eqref{eq:gh_inf_new} leads to a well-behaved source
(see, e.g., the dependence of $g$ in Eqs.~(2.26), (2.27), and (2.29)
of Ref.~\cite{Sasaki:1981sx}).

\section{Extreme Kerr limit}\label{sec:eK}

In the previous work~\cite{Nakamura:2016yjl}
for the analysis of the fundamental ($n=0$) QNM with ($\ell=2,\,m=2$)
in the extreme Kerr case, $q=a/M \to 1$,
we have derived a fitting curve of the peak location
in the Boyer--Lindquist coordinates as
\bea
 \frac{r_{\rm fit}}{M} = 1+1.4803 \,(-\ln q)^{0.503113} \,,
\label{eq:r_fit}
\eea
for the absolute value of the potential $|V_{\rm SN}|$
obtained by using the new $g$ presented in Eq.~\eqref{eq:new_g}
(called $V_{\rm NNT}$ in Ref.~\cite{Nakamura:2016yjl}).
In the WKB approximation, this peak location
is an important output obtained from the observation of the QNM GWs.

Here, we note that the event horizon radius is given by
\bea
 \frac{r_{+}}{M} &=& 1 + \sqrt{1-q^2}  
 \cr
 &=& 1 + \sqrt{2}\,(1-q)^{1/2}
 + O((1-q)^{3/2}) \,,
\eea
and the inner light ring radius~\cite{Bardeen:1972fi} is written as
\bea
 \frac{r_{\rm lr}}{M} &=& 2 + 2\cos\left[\frac{2}{3}
 \cos^{-1}\left(-q\right)\right] 
 \cr
 &=& 1 + \frac{2\sqrt{2}}{\sqrt{3}}\,(1-q)^{1/2}
 + O(1-q) \,.
\eea
The latter radius is evaluated in the equatorial ($\theta=\pi/2$) plane.
Although there are various studies on the relation between the QNMs
and the orbital frequency of the light ring orbit
(see a useful lecture note~\cite{Berti:2014bla}),
the peak location of the potential $r_{\rm fit}$
which derives the QNM frequencies, is much closer
to the horizon radius, $r_+/M \approx 1 + 1.414\,(1-q)^{1/2}$
than the inner light ring radius, $r_{\rm lr}/M \approx 1 + 1.633\,(1-q)^{1/2}$.

In Ref.~\cite{Nakamura:2016yjl},
to check the validity of $r_{\rm fit}$, 
we have evaluated the peak location (denoted by $r_{\rm p}$
in the Boyer--Lindquist coordinates)
semi-analytically by using a fitting formula for 
\bea
 \lambda = {}_sA_{\ell m} - 2am\omega + a^2\omega^2 \,,
\eea
with
\bea
 {}_{-2}A_{22} = 0.545652 + (6.02497+1.38591 \,i) (-\ln q)^{1/2} \,,
\eea 
which is a constant defined in Eq.~(25)
of Ref.~\cite{Berti:2009kk}.
Also, we have used the approximation for the ($n=0$) QNM frequency
with ($\ell=2,\,m=2$) in the extreme Kerr limit~\cite{Hod:2008zz},
\bea
 M \omega_{\rm ext} = \frac{M q}{r_+}-\frac{i}{4} \frac{r_+ - M}{r_+}
 \,.
\label{eq:Momega_e}
\eea
Then, defining $\epsilon$ by $q = 1-\epsilon^2$,
and expanding the potential $V_{\rm NNT}$
with respect to $\epsilon$, 
we derive the location $r_0$ of $dV_{\rm NNT}/dr^*=0$
instead of finding the peak location $r_{\rm p}$ of $|V_{\rm NNT}|$.
It is noted that the expression given in Eq.~\eqref{eq:Momega_e} can be considered
as the exact frequency derived by Leaver's method,
since we have discussed the extreme Kerr limit $\epsilon \to 0$.

In Appendix A of Ref.~\cite{Nakamura:2016gri},
we have found a good agreement between
the peak location of $|V_{\rm SN}|$
and the real part of the location of $dV_{\rm SN}/dr^*=0$.
The result was obtained as
\bea
 \frac{r_0}{M} = 1 + (1.44905-0.020157\,i) \,\epsilon \,,
\label{eq:r_0exp}
\eea
where the appearance of the $O(\epsilon^1)$ term
is consistent with the expression for $r_{\rm fit}$
given in Eq.~\eqref{eq:r_fit}
because $(-\ln q)^{1/2}=(1-q)^{1/2}+O((1-q)^{3/2})$.
Although it is consistent that both expressions, $r_{\rm fit}$ and $r_0$
have a correction of $O(\epsilon^1)$,
a different choice of $g$ from Eq.~\eqref{eq:new_g}
makes a difference in the coefficient of $O(\epsilon^1)$
in the estimation of $r_{\rm p}$ (and $r_0$).
In this section we study how robust the above estimation
of the peak location is.

We expand the event horizon radius as
\bea
 \frac{r_+}{M} = 1+\sqrt{2} \epsilon + O(\epsilon^3) \,, 
\eea
and the Boyer--Lindquist radial coordinate around $r_+$ as
\bea
 \frac{r}{M} &=& \frac{r_+}{M} + \xi \epsilon
 \cr
 &=& 1 + \sqrt{2} \epsilon + \xi \epsilon \,,
\eea
introducing a rescaled radial coordinate $\xi$
whose origin corresponds to the event horizon.
The tortoise coordinate is expressed as
\bea
 \frac{r^*}{M} = - \frac{\sqrt{2}}{2\epsilon} \ln \left(1+\frac{2\sqrt{2}}{\xi} \right)
 + O(\epsilon^0) \,.
\label{eq:rs_xi}
\eea
In the following analysis, 
we investigate the peak location
$r_{\rm p}/M=1 + \sqrt{2} \epsilon + \xi_{\rm p} \epsilon$,
keeping only the leading order with respect to $\epsilon$ for $\xi_{\rm p}$.
The QNM frequency in Eq.~\eqref{eq:Momega_e} is written as
\bea
 M \omega_{\rm ext} = 1 + \left(-\sqrt{2}-\frac{i}{4} \sqrt{2} \right) \epsilon
 + \left(1+\frac{i}{2} \right) \epsilon^2 \,,
\label{eq:Momega_exp}
\eea
up to $O(\epsilon^2)$.

The function $g$ in Eq.~\eqref{eq:new_g} is expanded for $\epsilon \ll 1$ as
\bea
g = \left( \frac{1}{4} \sqrt{2} +\frac{1}{4} \xi \right) \epsilon
+ O(\epsilon^2) \,.
\label{eq:new_g_exp}
\eea
In this expansion, the terms of $O(\xi^2)$
appear only at $O(\epsilon^2)$.
We focus on the leading-order modification of $O(\epsilon^1)$,
and consider a function linear in $\xi$.
Such a function is parametrized by two real
parameters $\mu$ and $\nu$ as
\bea
g &=& \epsilon+\frac{\sqrt{2}}{2} \left(\frac{1}{2}+\mu+i \nu\right)  \xi \epsilon \,.
\label{eq:linear_cal}
\eea
The function in Eq.~\eqref{eq:new_g_exp} is recovered when
$\mu=1/2$ and $\nu=0$,
except for the overall normalization of $g$ which does not contribute
to the potential
because of the dependence of $g$ in $P$ and $Q$
given by Eqs.~\eqref{eq:P} and \eqref{eq:Q}, respectively.
In the series expansion with respect to $\epsilon$, the potential in the
Sasaki--Nakamura equation (see Eq.~\eqref{eq:SN_V}) is formally written as
\bea
 V = 1 + \left( -2 \sqrt{2} -\frac{\sqrt{2}}{2} i
 \right) \epsilon + v^{(2)}_{\mu,\nu}(\xi) \epsilon^2 \,,
\label{eq:ptl_epsilon}
\eea
where we do not explicitly present the huge expression of $v^{(2)}_{\mu,\nu}(\xi)$.
It is noted that any $O(\epsilon^2)$ term in $g$ of Eq.~\eqref{eq:linear_cal}
does not contribute to the potential in the second order with respect to $\epsilon$.

Here, we define the error in the estimation of the QNM frequencies as
\bea
{\rm Err} &=& \sqrt{({\rm Err}_{\rm r})^2+({\rm Err}_{\rm i})^2} \,;
\cr
{\rm Err}_{\rm r} &=& \left|
\frac{{\rm Re}(\omega_{\rm WKB})-(1-\sqrt{2}\epsilon)}
{{\rm Re}(\omega_{\rm ext})-(1-\sqrt{2}\epsilon)} 
- 1 \right| 
\cr &\approx&
\left|
\frac{{\rm Re}(\omega_{\rm WKB})-(1-\sqrt{2}\epsilon)}
{\epsilon^2} - 1 \right| 
\,,
\cr
{\rm Err}_{\rm i} &\approx& \left|\frac{{\rm Im}(\omega_{\rm WKB})}{-\sqrt{2} \epsilon/4} - 1 \right|
\,,
\label{eq:errorE}
\eea
where we have calculated ${\rm Re}(\omega_{\rm ext})$ by using Eq.~\eqref{eq:Momega_exp},
and used the leading order of ${\rm Im}(\omega_{\rm ext})$
obtained from Eq.~\eqref{eq:Momega_exp} for ${\rm Err}_{\rm i}$.
Since the real part of the error, ${\rm Err}_{\rm r}$ is always tiny
in the case of small $\epsilon$
if we use $|{\rm Re}(\omega_{\rm WKB})/{\rm Re}(\omega_{\rm ext}) - 1|$
or $|({\rm Re}(\omega_{\rm WKB})-1)/({\rm Re}(\omega_{\rm ext})-1) - 1|$,
we have adopted the above estimator to normalize the error of the real part.
Note that this estimator is independent of $\epsilon$ in the limit $\epsilon \to 0$.

\begin{figure}[!t]
\begin{center}
 \includegraphics[width=0.49\textwidth,clip=true,bb=5 5 400 400]{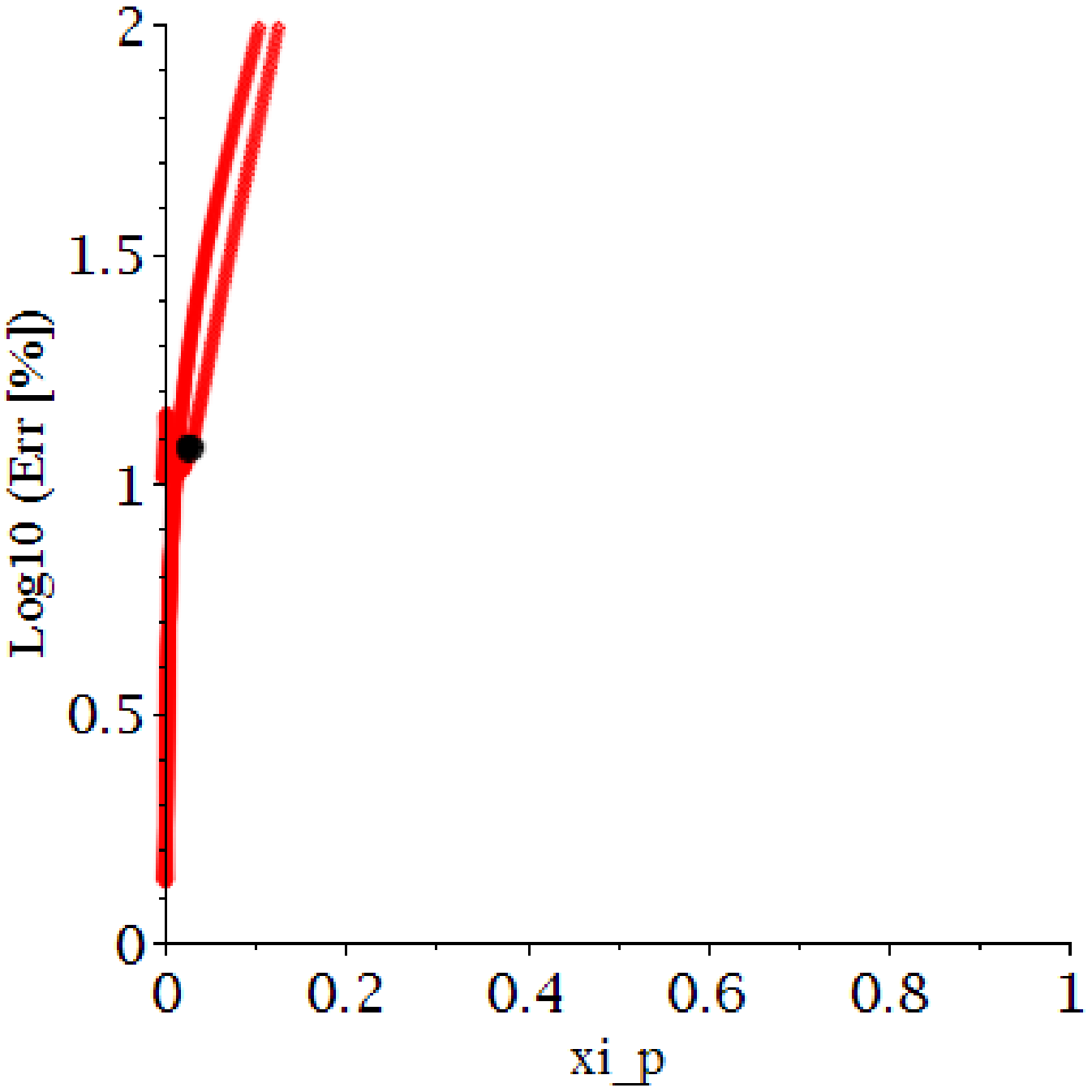}
 \includegraphics[width=0.49\textwidth,clip=true,bb=5 5 400 400]{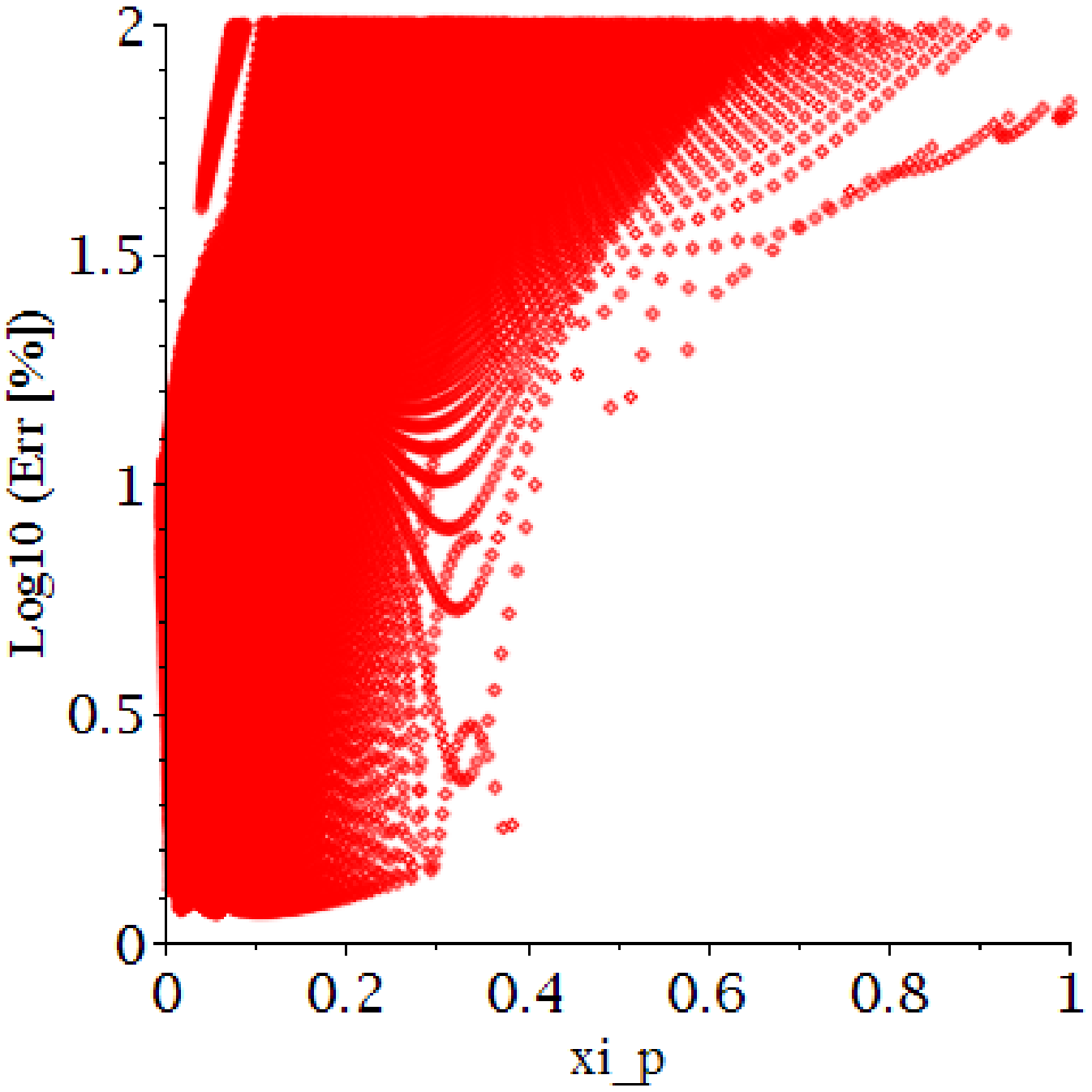}
\end{center}
 \caption{Scatter plots for the error estimation given
 by Eq.~\eqref{eq:errorE} with respect to $\xi_{\rm p}$,
 obtained by varying the parameters $\mu$ and $\nu$ in Eq.~\eqref{eq:linear_cal}.
 The left panel shows the case with fixed $\mu=1/2$, and the black dot denotes
 $\mu=1/2$ and $\nu=0$. In the right panel, $\mu$ is also variable.
 We do not find any point with a small error for $\xi_{\rm p}>0.4$.
 The empty region around $0<\xi_{\rm p}<0.1$ and $1.5<{\rm Log_{10}(Err)}<2.0$
 will be filled by the points if we use a much finer grid for the parameters $\mu$ and $\nu$.
}
 \label{fig:scatter}
\end{figure}

\begin{figure}[!t]
\begin{center}
 \includegraphics[width=0.49\textwidth,clip=true,bb=5 5 400 400]{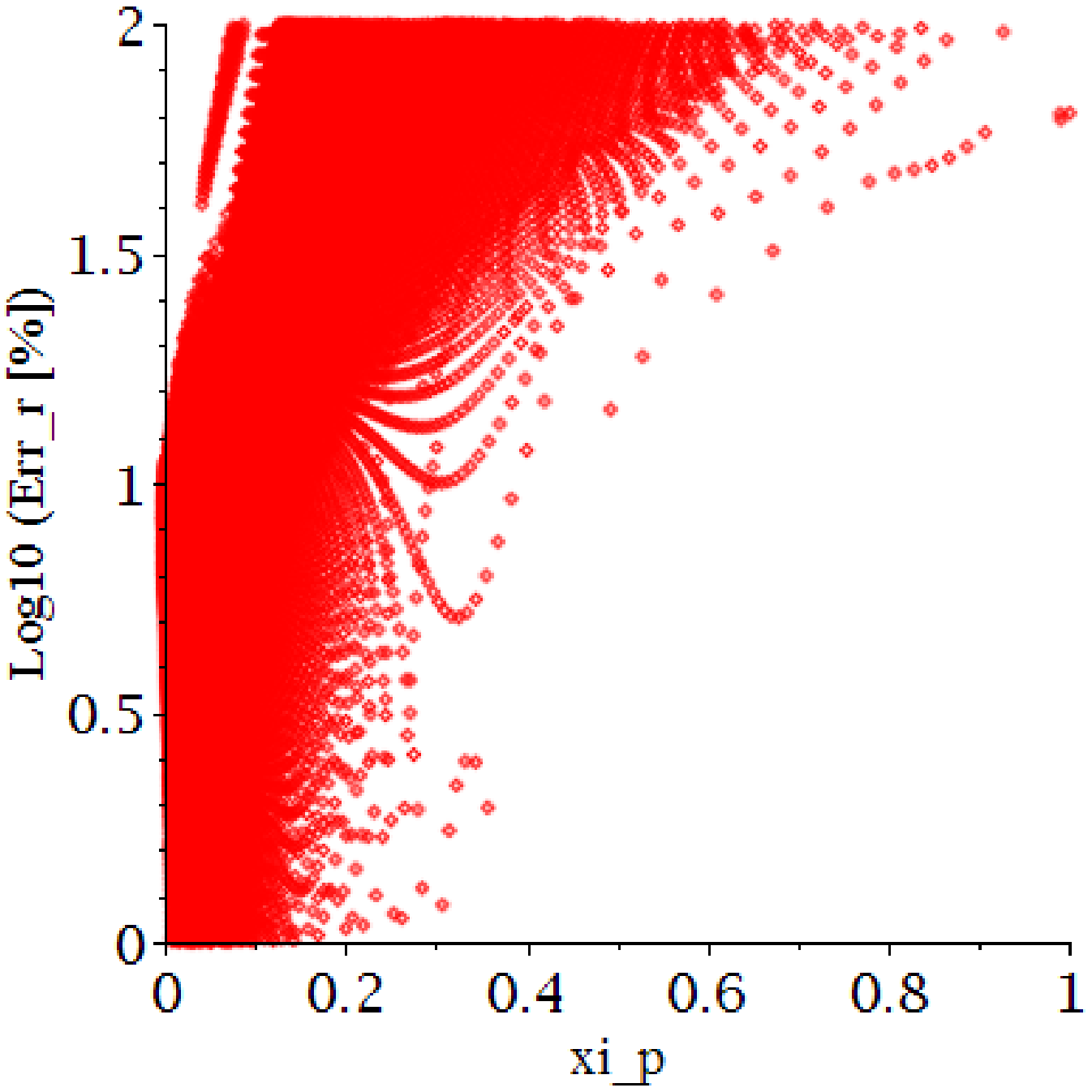}
 \includegraphics[width=0.49\textwidth,clip=true,bb=5 5 400 400]{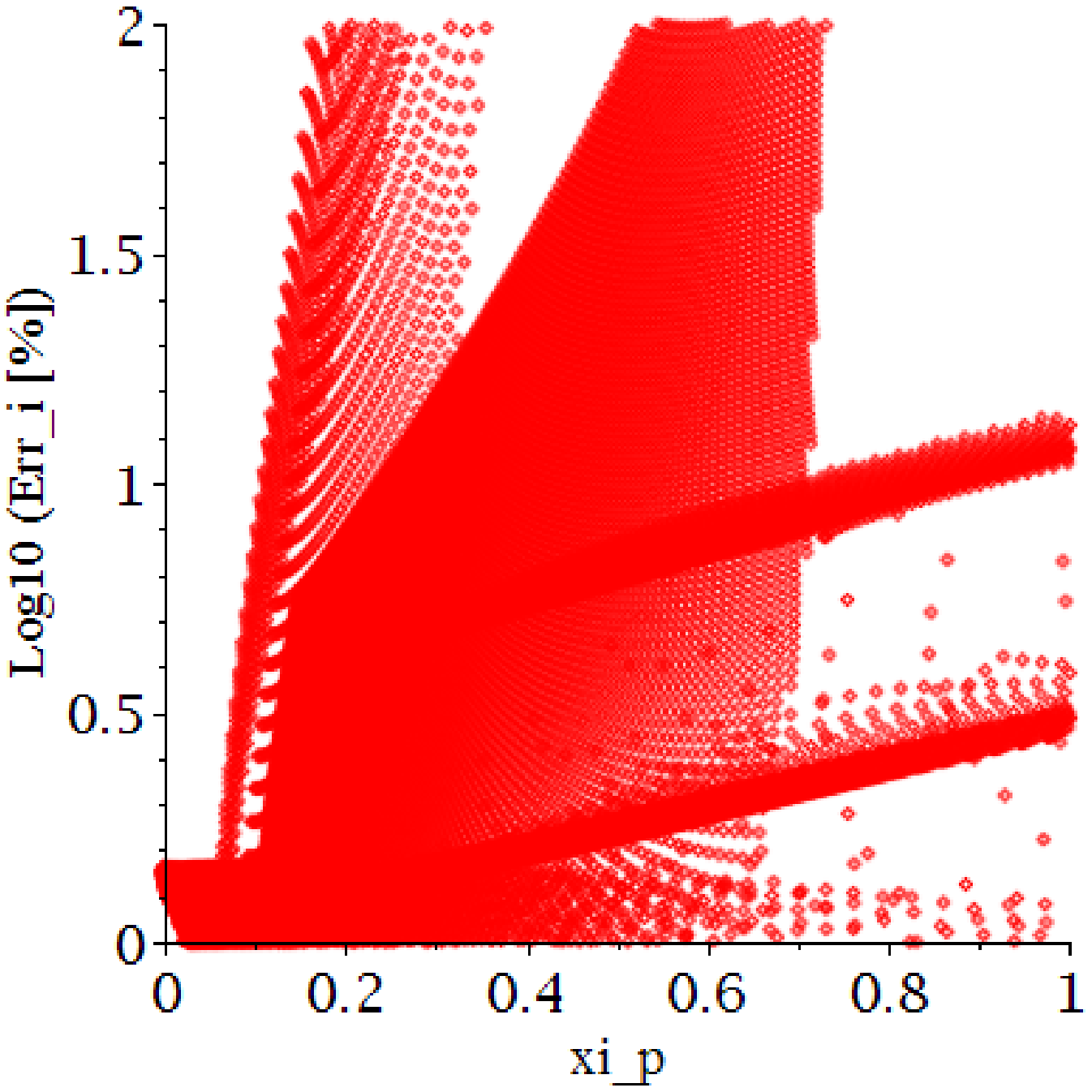}
\end{center}
 \caption{Scatter plots for ${\rm Err}_{\rm r}$ (left)
 and ${\rm Err}_{\rm i}$ (right) given
 by Eq.~\eqref{eq:errorE} with respect to $\xi_{\rm p}$,
 obtained by varying the parameters $\mu$ and $\nu$ in Eq.~\eqref{eq:linear_cal}.
}
 \label{fig:scatterRI}
\end{figure}

Varying the parameters $\mu$ and $\nu$,
we obtain Fig.~\ref{fig:scatter},
which shows the error in the estimation of the QNM frequencies
calculated by Eq.~\eqref{eq:errorE} with respect to $\xi_{\rm p}$
Figure~\ref{fig:scatterRI} shows ${\rm Err}_{\rm r}$ (the left panel)
and ${\rm Err}_{\rm i}$ (the right panel), respectively.
${\rm Err}$ is dominated by ${\rm Err}_{\rm r}$ for large $\xi_{\rm p}$.
We find from Fig.~\ref{fig:scatter}
that the region where ${\rm Err}$ given in Eq.~\eqref{eq:errorE}
is small spreads widely.
Although the minimum error of ${\rm Err} \approx 1.2\%$
is obtained for $\mu \approx -0.21$ and $\nu \approx 0.25$
in the analysis, 
there are many other combinations of $\mu$ and $\nu$
for which ${\rm Err}$ remains small,
and the region with small ${\rm Err}$ extends to a range of $0< \xi_{\rm p} < 0.4$.
Therefore, we should consider
that the peak of the potential is located in $0< \xi_{\rm p} < 0.4$.

In Refs.~\cite{Nakamura:2016gri,Nakano:2016sgf,Nakamura:2016yjl},
we have used the WKB analysis
to claim how deeply 
we can actually inspect the region close to the event horizon of a BH
by observing the QNM GWs.
Since the GWs cannot be localized 
and the QNMs are determined not only by the potential at the peak radius
but also by the curvature of the potential,
what we can claim is that the QNM frequency is determined by
the information ``around'' the peak of the potential.
Therefore, it is necessary to properly take into account this fact
in the interpretation of the estimated radius obtained
in Refs.~\cite{Nakamura:2016gri,Nakano:2016sgf,Nakamura:2016yjl}.

The uncertainty in the peak location can be discussed in the following manner.
Here, we use $r^*_0$ instead of $r^*_{\rm p}$ because $r^*_0$ is derived
easily in the analytical calculation.
Expanding Eq.~\eqref{eq:ptl_epsilon} with respect to
$\Delta r^*=r^*-r^*_0$ around $r^*_0$,
and using the QNM frequency in the WKB approximation of Eq.~\eqref{eq:WKB_QNM},
we have the radial wavenumber
[which corresponds to $W^{1/2}$ in Eq.~\eqref{eq:WKBfn}], as
\bea
k(r^*_0+\Delta r^*) &=& \sqrt{\omega_{\rm QNM}^2 - V}
\cr
&=& \sqrt{-\frac{i}{2}\sqrt{-2 \left. \frac{d^2V}{dr^{*2}} \right|_{r^*=r^*_0}}
-\frac{1}{2}\left. \frac{d^2V}{dr^{*2}} \right|_{r^*=r^*_0}(\Delta r^*)^2
+ \cdot\cdot\cdot } \,,
\label{eq:k}
\eea
where ($\cdot\cdot\cdot $) denotes the terms of higher order in the WKB approximation
or of $O((\Delta r^*)^3)$.
We note that $d^2V/dr^{*2}|_{r^*=r^*_0} = O(\epsilon^4)$
because $d/dr^* = O(\epsilon) \times d/d\xi$
and the $\xi$ dependence of $V$ is as given in Eq.~\eqref{eq:ptl_epsilon}.
If we expect that the uncertainty of the peak location is given
by the inverse of the wavenumber,
it may be estimated by the $\Delta r^*$ that solves
\bea
k(r^*_0+\Delta r^*) \Delta r^* = O(1) \,.
\label{eq:kdr}
\eea
Combining Eqs.~\eqref{eq:k} and \eqref{eq:kdr}, we derive 
\bea
\frac{\left|\Delta r^*\right|}{M} = O(\epsilon^{-1}) \,,
\label{eq:Drs}
\eea
which is translated into
the uncertainty in $\xi_0$ as
\bea
\left|\Delta \xi_0\right| = O(1) \,,
\eea
by using Eq.~\eqref{eq:rs_xi}.
This estimate is consistent with the extension of
the region where ${\rm Err}$ given in Eq.~\eqref{eq:errorE} 
is small in Fig.~\ref{fig:scatter}.

In our previous work~\cite{Nakamura:2016yjl},
we used only one potential which corresponds
to $\mu=1/2$ and $\nu=0$ in Eq.~\eqref{eq:linear_cal}.
The peak location was inside the light ring radius as shown
in Eqs.~\eqref{eq:r_fit} and \eqref{eq:r_0exp},
and we concluded that the QNM GWs were emitted ``around'' the peak location.
However, the meaning of the word ``around'' was not clear,
and Fig.~\ref{fig:scatter} gives a clear explanation
of it based on the error estimation
of the WKB frequencies compared with the exact QNM frequencies.
Using various potentials with the parameters, $\mu$ and $\nu$,
even if we change the threshold of the error estimator
\eqref{eq:errorE} from a few $\%$ to $10\%$,
the extension of the region does not change much from $\xi_{\rm p} \sim 0.4$.
Therefore, we conclude that the estimated peak location is restricted to
\bea
 \frac{r_{\rm p}}{M} \lesssim 1 + 1.8 \,(1-q)^{1/2} \,.
\label{eq:peakR}
\eea
The above result confirms that 
we can see the space-time sufficiently inside the ergoregion
($r_{\rm ergo} = 2M$ for the equatorial radius of the ergosurface)
and around the inner light ring $r_{\rm lr}/M \approx 1 + 1.633\,(1-q)^{1/2}$.

\section{Discussions}\label{sec:dis}

In the modification of the Sasaki--Nakamura equation,
we have found that the necessary and sufficient condition for
the fast fall-off at $r^* \to +\infty$
can be relaxed to the one given by Eq.~\eqref{eq:gh_inf_new},
if we assume that the free functions $g$ and $h$
can be expanded in a power series of $1/r$.
When we use $g$ in Eq.~\eqref{eq:new_g} which satisfies Eq.~\eqref{eq:gh_inf_new}, 
the potential is suitable for the WKB analysis of QNMs
(see Ref.~\cite{Nakamura:2016yjl}),
but the general expression of the potential is much more complicated
than that for $g$ in Eq.~\eqref{eq:orig_hg}.

One way to obtain a simple potential will be to keep $\gamma$ constant
in the Sasaki--Nakamura transformation.
$\gamma$ in Eq.~\eqref{eq:gamma} is rewritten as
\bea
\gamma = A^{2} 
\left( 1 - {\frac { \left( \lambda + 3\,iK'  \right)}{\Delta}}
\left(\frac{B}{A}\right)^2
- {\frac {i \left( 2\,K +i\Delta'  \right)}{\Delta}}
\left(\frac{B}{A}\right)
+ \left(\frac{B}{A}\right)' \right) \,.
\label{eq:gamma_comb}
\eea
Thus, it is not difficult to find $A$ and $B$ so that $\gamma$ is constant
because we may choose $B/A$ which leads $1/A^2$ for the expression
in the parenthesis of the above equation.
The difficult part arises from the condition for $A$ and $B$
that gives a short-ranged potential.
To derive such $A$ and $B$ is one of our future studies.

In the study of QNMs in the WKB method,
we have evaluated the uncertainty of the peak location of the potential
in the extreme Kerr limit.
This uncertainty is expressed as Eq.~\eqref{eq:peakR},
and is consistent with that expected from the equivalence principle.

Here, we should note that the imaginary parts of the QNM frequencies
become zero in the extreme Kerr limit,
and many overtones ($n \neq 0$) accumulate
at one frequency (see, e.g., Fig.~3 in Ref.~\cite{Cook:2014cta}
and a recent work~\cite{Richartz:2015saa}).
Therefore, observing QNM GWs in the near-extremal Kerr case
will be very different from the other case,
and further studies are required to extract the information from extreme Kerr BHs.

Finally, thanks to the recent GW observation, GW150914, 
we have entered the next stage of using GWs
to extract new physics.
To test the strong gravitational field around BHs,
the QNMs are simple and useful,
and the QNM GWs are the target 
not only for the second-generation GW detectors such as
Advanced LIGO (aLIGO)~\cite{TheLIGOScientific:2014jea}, 
Advanced Virgo (AdV)~\cite{TheVirgo:2014hva}, 
and KAGRA~\cite{Somiya:2011np,Aso:2013eba},
but also for space-based GW detectors
such as eLISA~\cite{Seoane:2013qna} and DECIGO~\cite{Seto:2001qf}.
The enhancement of the signal-to-noise ratio
by the third-generation detectors such as the Einstein Telescope
(ET)~\cite{Punturo:2010zz} will significantly improve
the precision of the test of general relativity.

\section*{Acknowledgments}

This work was supported by MEXT 
Grant-in-Aid for Scientific Research on Innovative Areas,
``New Developments in Astrophysics Through Multi-Messenger Observations
of Gravitational Wave Sources,'' Nos.~24103001 and 24103006 (HN, TT, TN),
JSPS Grant-in-Aid for Scientific Research (C), No.~16K05347 (HN),
JSPS Grant-in-Aid for Young Scientists (B), No.~25800154 (NS),
and Grant-in-Aid from the Ministry of Education, Culture, Sports,
Science and Technology (MEXT) of Japan No.~15H02087 (TT, TN).

\appendix

\section{Quasinormal mode in the WKB approximation}\label{sec:SNw}

We schematically write Eq.~\eqref{eq:finalSN} as
\bea
 \frac{d^2\psi}{dr^{*2}} + W(r^*) \psi =0 \,,
\label{eq:scheW}
\eea
where $W = \omega^2-V_{\rm SN}$.
In Ref.~\cite{Schutz:1985zz}, the QNM frequencies
are discussed as a ``second-order turning point'' problem
in the WKB approximation (see also Ref.~\cite{BenderOrszag}).
We prepare two WKB solutions,
\bea
 \psi_{1}^{\rm WKB} \approx [W(r^*)]^{-1/4}
 \exp \left( \pm i \int_{r^*_2}^{r^*} [W(x)]^{1/2} dx \right) \,,
 \cr
 \psi_{2}^{\rm WKB} \approx [W(r^*)]^{-1/4}
 \exp \left( \pm i \int_{r^*}^{r^*_1} [W(x)]^{1/2} dx \right) \,,
\label{eq:WKBfn}
\eea
where $r^*_1$ and $r^*_2$ are the turning points,
and also parabolic cylinder functions for $r^*_1 < r^* < r^*_2$
(see Eq.~(5) of Ref.~\cite{Schutz:1985zz}).
The QNM frequencies are derived in the matching condition
for the outgoing (from the peak location of the potential)
solutions of $\psi_{1}^{\rm WKB}$ and $\psi_{2}^{\rm WKB}$.
This means that we choose the signs in Eq.~\eqref{eq:WKBfn} appropriately.

As a usual picture, the peak location is calculated by
\bea
 \frac{dW(r^*)}{dr^*} = 0 \,,
\eea
the solution is denoted by $r^*_0$, and $r^*_1 < r^*_0 < r^*_2$
in the case of a real potential. We extend this to a complex potential.
Therefore, $r^*_0$ is in the complex plane.



\begin{thebibliography}{DUM}

\bibitem{Abbott:2016blz} 
  B.~P.~Abbott {\it et al.} [LIGO Scientific and Virgo Collaborations],
  Phys.\ Rev.\ Lett.\  {\bf 116}, 061102 (2016)
  [arXiv:1602.03837 [gr-qc]].

\bibitem{TheLIGOScientific:2016htt} 
  B.~P.~Abbott {\it et al.} [LIGO Scientific and Virgo Collaborations],
  Astrophys.\ J.\  {\bf 818}, L22 (2016)
  [arXiv:1602.03846 [astro-ph.HE]].

\bibitem{Kinugawa:2014zha} 
  T.~Kinugawa, K.~Inayoshi, K.~Hotokezaka, D.~Nakauchi and T.~Nakamura,
  Mon.\ Not.\ Roy.\ Astron.\ Soc.\  {\bf 442}, 2963 (2014)
  [arXiv:1402.6672 [astro-ph.HE]].

\bibitem{Kinugawa:2015nla} 
  T.~Kinugawa, A.~Miyamoto, N.~Kanda and T.~Nakamura,
   Mon.\ Not.\ Roy.\ Astron.\ Soc.\  {\bf 456}, 1093 (2016)
   [arXiv:1505.06962 [astro-ph.SR]].

\bibitem{Kinugawa:2016mfs} 
  T.~Kinugawa, H.~Nakano and T.~Nakamura,
  Prog. Theor. Exp. Phys. (2016), 031E01
  [arXiv:1601.07217 [astro-ph.HE]].

\bibitem{Hartwig:2016nde} 
  T.~Hartwig, M.~Volonteri, V.~Bromm, R.~S.~Klessen, E.~Barausse, M.~Magg and A.~Stacy,
  arXiv:1603.05655 [astro-ph.GA].

\bibitem{TheLIGOScientific:2016wfe} 
  B.~P.~Abbott {\it et al.} [LIGO Scientific and Virgo Collaborations],
  arXiv:1602.03840 [gr-qc].

\bibitem{Healy:2014yta} 
  J.~Healy, C.~O.~Lousto and Y.~Zlochower,
  Phys.\ Rev.\ D {\bf 90}, 104004 (2014)
  [arXiv:1406.7295 [gr-qc]].

\bibitem{Ghosh:2015jra} 
  A.~Ghosh, W.~Del Pozzo and P.~Ajith,
  arXiv:1505.05607 [gr-qc].

\bibitem{Berti:2007zu} 
  E.~Berti, J.~Cardoso, V.~Cardoso and M.~Cavaglia,
  Phys.\ Rev.\ D {\bf 76}, 104044 (2007)
  [arXiv:0707.1202 [gr-qc]].

\bibitem{Nakano:2015uja} 
  H.~Nakano, T.~Tanaka and T.~Nakamura,
  Phys.\ Rev.\ D {\bf 92}, 064003 (2015)
  [arXiv:1506.00560 [astro-ph.HE]].

\bibitem{Konoplya:2016pmh} 
  R.~Konoplya and A.~Zhidenko,
  Phys.\ Lett.\ B {\bf 756}, 350 (2016)
  [arXiv:1602.04738 [gr-qc]].

\bibitem{Yunes:2016jcc} 
  N.~Yunes, K.~Yagi and F.~Pretorius,
  arXiv:1603.08955 [gr-qc].

\bibitem{TheLIGOScientific:2016src} 
  B.~P.~Abbott {\it et al.} [LIGO Scientific and Virgo Collaborations],
  arXiv:1602.03841 [gr-qc].

\bibitem{Kerr:1963ud} 
  R.~P.~Kerr,
  Phys.\ Rev.\ Lett.\  {\bf 11}, 237 (1963).

\bibitem{Teukolsky:1973ha} 
  S.~A.~Teukolsky,
  Astrophys.\ J.\  {\bf 185}, 635 (1973).

\bibitem{Chandrasekhar:1976zz} 
  S.~Chandrasekhar and S.~L.~Detweiler,
  Proc.\ Roy.\ Soc.\ Lond.\ A {\bf 350}, 165 (1976).

\bibitem{Nakamura:2016gri} 
  T.~Nakamura, H.~Nakano and T.~Tanaka,
  Phys.\ Rev.\ D {\bf 93}, 044048 (2016)
  [arXiv:1601.00356 [astro-ph.HE]].

\bibitem{Nakano:2016sgf} 
  H.~Nakano, T.~Nakamura and T.~Tanaka,
  Prog. Theor. Exp. Phys. (2016) 031E02
  [arXiv:1602.02875 [gr-qc]].

\bibitem{Detweiler:1977gy} 
  S.~L.~Detweiler,
  Proc.\ Roy.\ Soc.\ Lond.\ A {\bf 352}, 381 (1977).

\bibitem{Mashhoon:1985cya} 
  B.~Mashhoon,
  Phys.\ Rev.\ D {\bf 31}, 290 (1985).

\bibitem{Schutz:1985zz} 
  B.~F.~Schutz and C.~M.~Will,
  Astrophys.\ J.\  {\bf 291}, L33 (1985).

\bibitem{Sasaki:1981kj} 
  M.~Sasaki and T.~Nakamura,
  Phys.\ Lett.\ A {\bf 89}, 68 (1982).

\bibitem{Sasaki:1981sx} 
  M.~Sasaki and T.~Nakamura,
  Prog.\ Theor.\ Phys.\  {\bf 67}, 1788 (1982).

\bibitem{Nakamura:1981kk} 
  T.~Nakamura and M.~Sasaki,
  Phys.\ Lett.\ A {\bf 89}, 185 (1982).

\bibitem{Hughes:2000pf} 
  S.~A.~Hughes,
  Phys.\ Rev.\ D {\bf 62}, 044029 (2000)
  [Phys.\ Rev.\ D {\bf 67}, 089902 (2003)]
  [gr-qc/0002043].

\bibitem{Nakamura:2016yjl} 
  T.~Nakamura and H.~Nakano,
  Prog. Theor. Exp. Phys. (2016) 041E01
  [arXiv:1602.02385 [gr-qc]].

\bibitem{Berti:2005ys} 
  E.~Berti, V.~Cardoso and C.~M.~Will,
  Phys.\ Rev.\ D {\bf 73}, 064030 (2006)
  [gr-qc/0512160].

\bibitem{Leaver:1985ax} 
  E.~W.~Leaver,
  Proc.\ Roy.\ Soc.\ Lond.\ A {\bf 402}, 285 (1985).

\bibitem{Regge:1957td}
  T.~Regge and J. A. Wheeler,
  Phys.\ Rev.\  {\bf 108}, 1063 (1957).

\bibitem{Bardeen:1972fi} 
  J.~M.~Bardeen, W.~H.~Press and S.~A.~Teukolsky,
  Astrophys.\ J.\  {\bf 178}, 347 (1972).

\bibitem{Berti:2014bla} 
  E.~Berti,
  arXiv:1410.4481 [gr-qc].

\bibitem{Berti:2009kk} 
  E.~Berti, V.~Cardoso and A.~O.~Starinets,
  Class.\ Quant.\ Grav.\  {\bf 26}, 163001 (2009)
  [arXiv:0905.2975 [gr-qc]].

\bibitem{Hod:2008zz} 
  S.~Hod,
  Phys.\ Rev.\ D {\bf 78}, 084035 (2008)
  [arXiv:0811.3806 [gr-qc]].

\bibitem{Cook:2014cta} 
  G.~B.~Cook and M.~Zalutskiy,
  Phys.\ Rev.\ D {\bf 90}, 124021 (2014)
  [arXiv:1410.7698 [gr-qc]].

\bibitem{Richartz:2015saa} 
  M.~Richartz,
  Phys.\ Rev.\ D {\bf 93}, 064062 (2016)
  [arXiv:1509.04260 [gr-qc]].

\bibitem{TheLIGOScientific:2014jea} 
  J.~Aasi {\it et al.}  [LIGO Scientific Collaboration],
  Class.\ Quant.\ Grav.\  {\bf 32}, 074001 (2015)
  [arXiv:1411.4547 [gr-qc]].

\bibitem{TheVirgo:2014hva} 
  F.~Acernese {\it et al.}  [VIRGO Collaboration],
  Class.\ Quant.\ Grav.\  {\bf 32}, 024001 (2015)
  [arXiv:1408.3978 [gr-qc]].

\bibitem{Somiya:2011np} 
  K.~Somiya [KAGRA Collaboration],
  Class.\ Quant.\ Grav.\  {\bf 29}, 124007 (2012)
  [arXiv:1111.7185 [gr-qc]].

\bibitem{Aso:2013eba} 
  Y.~Aso {\it et al.}  [KAGRA Collaboration],
  Phys.\ Rev.\ D {\bf 88}, 043007 (2013)
  [arXiv:1306.6747 [gr-qc]].

\bibitem{Seoane:2013qna} 
  P.~A.~Seoane {\it et al.} [eLISA Collaboration],
  arXiv:1305.5720 [astro-ph.CO].

\bibitem{Seto:2001qf} 
  N.~Seto, S.~Kawamura and T.~Nakamura,
  Phys.\ Rev.\ Lett.\  {\bf 87}, 221103 (2001)
  [astro-ph/0108011].
 
\bibitem{Punturo:2010zz} 
  M.~Punturo {\it et al.},
  Class.\ Quant.\ Grav.\  {\bf 27}, 194002 (2010).

\bibitem{BenderOrszag} 
  C.~M.~Bender and S.~A.~Orszag,
  {\it Advanced Mathematical Methods for Scientists and Engineers 1, Asymptotic Methods and Perturbation Theory}
  (Springer, New York, 1999).
 
\end{thebibliography}
\end{document}